\documentclass[amsmath,amssymb,reprint,groupedaddress, showpacs,aps,prl]{revtex4-1}
\usepackage{latexsym}
\usepackage{amssymb}
\usepackage{amsmath}
\usepackage{amsfonts}
\usepackage{color}
\usepackage{hyperref}
\usepackage{mathtools}
\usepackage{tikz-cd} 
\usepackage[letterpaper,left=1in,top=1in,bottom=1in,right=1in,includeheadfoot,headheight=.6pt]{geometry}

\begin{document}
\title{Nash embedding and equilibrium in pure quantum states}
\author{Faisal Shah Khan}
\affiliation{Quantum Computing Research Group, Applied Math \& Sciences, Khalifa University, Abu Dhabi UAE}
\email{faisal.khan@kustar.ac.ae}
\author{Travis S. Humble}
\affiliation{Quantum Computing Institute, Oak Ridge National Laboratory, Oak Ridge, Tennessee USA}
\email{humblets@ornl.gov}

\begin{abstract}
With respect to probabilistic mixtures of the strategies in non-cooperative games, quantum game theory provides guarantee of fixed-point stability, the so-called Nash equilibrium. This permits players to choose mixed quantum strategies that prepare mixed quantum states optimally under constraints. We show here that fixed-point stability of Nash equilibrium can also be guaranteed for pure quantum strategies via an application of the Nash embedding theorem, permitting players to prepare pure quantum states optimally under constraints. 
\end{abstract}

\pacs{42.50.Ar,03.67.Bg,42.79.Sz}
\maketitle 

\section{Introduction}

As quantum technologies increase in scale and complexity, constrained optimization of the underlying quantum processes will also increase in importance. A good example is the case of a quantum Internet as a network of quantum devices that process and relay quantum information using teleportation and entanglement swapping. Connectivity constraints arise naturally when optimizing classical network resources \cite{Fabrikant, Albers}, and similar constraints manifest in the design and operation of quantum networks \cite{Scarpa, Liu, Venkat}. Similarly, quantum computers implementing fault-tolerant operations must optimize circuit design in order to minimize the decoherence due to the intrinsic physical noise in gate operation while also adhering to constraints in the circuit layout, scheduling, and parallelization. Finding optimal solutions to such constrained optimization problems are expected to maximize the efficiency and performance of quantum information technologies.
\par
Non-cooperative game theory \cite{Binmore} studies optimization under constraints and can offer useful insights into the engineering of scalable quantum technologies such as optimal bounds on errors and their correction in quantum computations. Any quantum physical process modeled as a non-cooperative game describes a {\it quantum game} of the same type. The first instance of non-cooperative game-theoretic modeling of quantum physical processes appears to be the 1980 work of A. Blaquiere \cite{Blaquier}, where wave mechanics are considered as a two player, zero-sum (strictly competitive) differential game and a mini-max result is established for certain quantum physical aspects. The more recent and more sustained game-theoretic treatment of quantum physical processes was initiated in 1999 with the work of Meyer \cite{Meyer}. Meyer's work considered quantum computational and quantum algorithmic aspects of quantum physics as non-cooperative games. 
\par
The year 1999 also saw the publication of the paper \cite{Eisert} by Eisert et al. in which a quantum informational model for the informational component of two players games was considered. This consideration was in the same spirit as the consideration of randomizing in a game which produces the so-called {\it mixed} game played with mixed strategies. The quantum informational model of Eisert et al. produces a {\it quantized} game. The inspiration for considering extensions of the informational aspect of games to larger domains comes from John Nash's famous theorem \cite{Nash} in economics which not only innovates the solution concept of non-cooperative games as an equilibrium problem but, for probabilistic extensions of finite non-cooperative games, also guarantees its existence. 
\par
The promise of a Nash equilibrium solution is a foundational concept for game theory as it may be used to guarantee the behavior for the non-cooperating players. The relative simplicity of the proof of Nash's theorem for the existence of an equilibrium in mixed strategies in conventional games relies entirely on Kakutani's fixed-point theorem \cite{Kakutani}. For quantum games, Meyer established the existence of Nash equilibrium in mixed strategies, which are modeled as mixed quantum states, using Glicksberg's \cite{Glick} extension of Kakutani's fixed point theorem to topological vector spaces.
\par
In this contribution, we note that the Kakutani fixed-point theorem does not apply directly to quantum games played with pure quantum strategies. But, one can use Nash's embedding of compact Riemannian manifolds into Euclidean space \cite{Nash1} (Nash's other, mathematically more famous theorem) and, under appropriate conditions, indirectly apply the Kakutani fixed-point theorem to guarantee Nash equilibrium in pure quantum strategies. We begin with a mathematically formal discussion of non-cooperative game theory and fixed-points.  

\section{Non-cooperative games and Nash equilibrium}\label{Nash}

An $N$ player, non-cooperative game in normal form is a function $\Gamma$ 
\begin{equation}\label{Game}
\Gamma: \prod_{i=1}^N S_i \longrightarrow O,
\end{equation}
with the additional feature of the notion of non-identical preferences over the elements of the set of {\it outcomes} $O$, for every ``player'' of the game. The preferences are a pre-ordering of the elements of $O$, that is, for $l,m,n \in O$
\begin{equation}
m \preceq m,  \hspace{2mm}  {\rm and}  \hspace{2mm}  l \preceq m \hspace{2mm} {\rm and} \hspace{2mm}  m \preceq n \implies  l \preceq n.
\end{equation} 
where the symbol $\preceq$ denotes ``of less or equal preference''. Preferences are typically quantified numerically for the ease of calculation of the payoffs. To this end, functions $\Gamma_i$ are introduced which act as the {\it payoff function} for each player $i$ magenta and typically map elements of $O$ into the real numbers in a way that preserves the preferences of the players. That is, $\preceq$ is replaced with $\leq$ when analyzing the payoffs. The factor $S_i$ in the domain of $\Gamma$ is said to be the {\it strategy set} of player $i$, and a {\it play} of $\Gamma$ is an $n$-tuple of strategies, one per player, producing a payoff to each player in terms of his preferences over the elements of $O$ in the image of $\Gamma$.
\par
A non-cooperative $N$-player quantum game in normal form arises from (\ref{Game}) when one introduces quantum physically relevant restrictions. We declare a pure (strategy) quantum game to be any unitary function 
\begin{equation}\label{quantgame}
Q: \otimes_{i=1}^N \mathbb{C}P^{d_i} \longrightarrow \otimes_{i=1}^{N}\mathbb{C}P^{d_i}
\end{equation}
where $\mathbb{C}P^{d_i}$ is the $d_i$-dimensional complex projective Hilbert space of pure quantum states. The latter are typically referred to as $d$-ary ``quantum digits" or {\it qudits}. By analogy with mixed game extensions, where players' strategies are probability distributions over the elements of some set, the strategies of each player in a quantum game consist of quantum superpositions over the elements of a set of observable states in $\mathbb{C}P^{d_i}$. These strategic choices are then mapped by $Q$ into $\otimes_{i=1}^{N} \mathbb{C}P^{d_i}$, over the elements of which the players have non-identical preferences defined using the overlap of two qudits as the payoff functions. 
\par
The overlap of two qudits is a complex number in general. This is in contrast to the more standard practice in classical game theory of defining payoff functions that map into the real numbers. Indeed, in the current context of pure strategy quantum games, the expected value of an observable computed after quantum measurement can be taken as the payoff function mapping into the set of real numbers. However, as we will show later in detail, the non-linearity of the expected value of an observable fails to guarantee Nash equilibrium whereas the linearity of the overlap does not.
\par
{\it Nash equilibrium} is a play of $\Gamma$ in which every player employs a strategy that is a best reply, with respects to his preferences over the outcomes, to the strategic choice of every other player. In other words, unilateral deviation from a Nash equilibrium by any one player in the form of a different choice of strategy will produce an outcome which is less preferred by that player than before. Following Nash, we say that a play $p'$  of $\Gamma$ {\it counters} another play $p$ if $\Gamma_i(p') \geq \Gamma_i(p)$ for all players $i$, and that a self-countering play is an (Nash) equilibrium. 
\par
Let $C_{p}$ denote the set of all the plays of $\Gamma$ that counter $p$. Denote $\prod_{i=1}^N S_i$ by $S$ for notational convenience, and note that $C_{p} \subset S$ and therefore $C_{p} \in 2^S$. Further note that the game $\Gamma$ can be factored as  
\begin{equation}\label{factor}
\Gamma:  S \xrightarrow{\Gamma_C} 2^S \xrightarrow{E} O
\end{equation}
where to any play $p$ the map $\Gamma_C$ associates its countering set $C_p$ via the payoff functions $\Gamma_i$. The set-valued map $\Gamma_C$ may be viewed as a preprocessing stage where players seek out a self-countering play, and if one is found, it is mapped to its corresponding outcome in $O$ by the function $E$. The condition for the existence of a self-countering play, and therefore of a Nash equilibrium, is that $\Gamma_C$ have a fixed point, that is, an element $p^* \in S$ such that $p^* \in \Gamma_C(p^*)=C_{p^*}$.
\par
In a general set-theoretic setting for non-cooperative games, the map $\Gamma_C$ may not have a fixed point. Hence, not all non-cooperative games will have a Nash equilibrium. However, according to Nash's theorem, when the $S_i$ are finite and the game is extended to its {\it mixed} version, that is, the version in which randomization via probability distributions is allowed over the elements of all the $S_i$, as well as over the elements of $O$, then $\Gamma_C$ has at least one fixed point and therefore at least one Nash equilibrium. 
\par
Formally, given a game $\Gamma$ with finite $S_i$ for all $i$, its mixed version is the product function 
\begin{equation}\label{mixedgame} 
\Lambda: \prod_{i=1}^N \Delta(S_i) \longrightarrow \Delta(O)
\end{equation}
where $\Delta(S_i)$ is the set of probability distributions over the $i^{\rm{th}}$ player's strategy set $S_i$, and the set $\Delta(O)$ is the set of probability distributions over the outcomes $O$. Payoffs are now calculated as {\it expected payoffs}, that is, weighted averages of the values of $\Gamma_i$, for each player $i$, with respect to probability distributions in $\Delta(O)$ that arise as the product of the plays of $\Lambda$. Denote the expected payoff to player $i$ by the function $\Lambda_i$. Also, note that $\Lambda$ restricts to $\Gamma$.
In these games, at least one Nash equilibrium play is guaranteed to exist as a fixed point of $\Lambda$ via Kakutani's fixed-point theorem.

\vspace{5mm}
\noindent {\bf Kakutani fixed-point theorem}: {\it Let $S \subset \mathbb{R}^n$ be nonempty, compact, and convex, and let $F: S\rightarrow 2^S$ be an upper semi-continuous set-valued mapping such that $F(s)$ is non-empty, closed, and convex for all $s \in S$. Then there exists some $s^* \in S$ such that $s^* \in F(s^*)$}.
\vspace{3mm}

To see this, make $S=\prod_{i=1}^N \Delta(S_i)$. Then $S \subset \mathbb{R}^n$ and $S$ is non-empty, bounded, and closed because it is a finite product of finite non-empty sets. The set $S$ is also convex because its the convex hull of the elements of a finite set. Next, let $C_p$ be the set of all plays of $\Lambda$ that counter the play $p$. Then $C_p$ is non-empty, closed, and convex. Further, $C_p \subset S$ and therefore $C_p \in 2^S$. Since $\Lambda$ is a game, it factors according to (\ref{factor})
\begin{equation}
\Lambda: S \xrightarrow{\Lambda_C} 2^S \xrightarrow{E_{\Pi}} \Delta(O)
\end{equation}
where the map $\Lambda_C$ associates a play to its countering set via the payoff functions $\Lambda_i$. Since $\Lambda_i$ are all continuous, $\Lambda_C$ is continuous. Further, $\Lambda_C(s)$ is non-empty, closed, and convex for all $s \in S$ (we will establish the convexity of $\Lambda_C(s)$ below; the remaining conditions are also straightforward to establish). Hence, Kakutani's theorem applies and there exists an $s^* \in S$ that counters itself, that is, $s^* \in \Lambda_C(s^*)$, and is therefore a Nash equilibrium. The function $E_{\Pi}$ simply maps $s^*$ to $\Delta(O)$ as the product probability distribution from which the Nash equilibrium expected payoff is computed for each player. 
\par
The convexity of the $\Lambda_C(s)=C_p$ is straight forward to show. Let $r, s \in C_p$. Then
\begin{equation}\label{counteringeq}
\Lambda_i(r) \geq \Lambda_i(p) \quad {\rm and} \quad \Lambda_i(s) \geq \Lambda_i(p)
\end{equation} 
for all $i$. Now let $0 \leq \mu \leq 1$ and consider the convex combination $\mu r + (1-\mu)s$ which we will show to be in $C_p$. First note that $\mu r + (1-\mu)s \in S$ because $S$ is the product of the convex sets $\Delta(S_i)$. Next, since the $\Lambda_i$ are all linear, and because of the inequalities in (\ref{counteringeq}) and the restrictions on the values of $\mu$,
\begin{equation}\label{linearity}
\Lambda_i(\mu r+ (1- \mu)s)=\mu \Lambda_i(r) + (1- \mu) \Lambda_i(s) \geq \Lambda_i(p)
\end{equation}
whereby $\mu r + (1-\mu)s \in C_p$ and $C_p$ is convex. 
\par
Going back to the game $\Gamma$ in (\ref{Game}) defined in the general set-theoretic setting, Kakutani's theorem would apply to $\Gamma$ if the conditions are right, that is, whenever the image set of $\Gamma$ is pre-ordered and $\Gamma_i$ is both linear and preserves the pre-order.
\par
Kakutani's fixed-point theorem can be generalized to include subsets $S$ of convex topological vector spaces, as was done by Glicksberg in \cite{Glick}. 
%
\noindent Using Glicksberg's fixed-point theorem, one can show that Nash equilibrium exists in games where the strategy sets are infinite or possibly even uncountably infinite. As mentioned earlier in the Introduction, Meyer used Glicksberg's fixed-point theorem to establish the guarantee of Nash equilibrium in mixed quantum games. This is not surprising given that probabilistic mixtures form a convex structure, which is an essential ingredient for fixed-point theorems to hold on ``flat'' manifolds such as $\mathbb{R}^m$.

\section{Pure Quantum Games and Nash equilibrium}\label{pureqgames}
No fixed-point theorem guarantee for Nash equilibrium in pure quantum strategies is known to exist in the literature. This is surprising perhaps given the rich, albeit non-convex, mathematical structure of $\mathbb{C}P^{n}$. More precisely, $\mathbb{C}P^{n}$ has a compact Riemannian (Kahler in fact) manifold structure with positive sectional curvature with respect to the Fubini-Study metric \cite{Bengtsson}. We use the richness of the mathematical structure of $\mathbb{C}P^{n}$ here to produce a guarantee of Nash equilibrium in pure quantum strategies, under restrictive conditions, by invoking Nash's embedding theorem:

\vspace{5mm}
\noindent {\bf Nash embedding theorem}: {\it For every compact Riemannian manifold $M$,  there exists an isometric embedding of $M$ into $\mathbb{R}^m$ for a suitably large $m$}. 
\vspace{3mm}

The Nash embedding theorem tells us that $\mathbb{C}P^{n}$ is diffeomorphic to its image under a length preserving map into $\mathbb{R}^m$. The homeomorphism underlying this diffeomorphism allows us to treat $\mathbb{C}P^{n}$ and its image inside $\mathbb{R}^m$ as topologically equivalent; hence, we can treat $\mathbb{C}P^{n}$ as a sub-manifold $S$ of $\mathbb{R}^m$ and look for a fixed-point guarantee for continuous set-valued functions $F$ 
\begin{equation}\label{eqn::embed}
\mathbb{C}P^n \xhookrightarrow{e} S \xrightarrow{F} 2^S
\end{equation}
via Kakutani's fixed-point theorem, where $e$ is the Nash embedding. To this end, recall that the Kakutani fixed-point theorem requires $S$ to be compact and convex. Since $e$ is a homeomorphism and $\mathbb{C}P^{n}$ is compact, $S$ is compact in $\mathbb{R}^m$. However, $S$ is not necessarily convex as homeomorphisms do not preserve convexity in general.
Note also that the linearity of Nash embedding $e$ would be insufficient to ensure convexity of $S$, for an element $q$ of $\mathbb{C}P^n$ is equivalent to all scalar multiples $\lambda q$, for $\lambda \neq 0$. This violates the definition of convexity, which requires the possibility that $\lambda = 0$. 
\par
But suppose for the moment that there exists a convex embedding $S$ of $\mathbb{C}P^{n}$ into $\mathbb{R}^m$. For the Kakutani fixed-point theorem to be applicable to $F$, one encases $S$ in a simplex $\Delta$, establishes Kakutani's theorem on $\Delta$ via barycentric subdivision and the Brouwer fixed-point theorem \cite{Brouwer} , and then constructs a {\it retract} function 
\begin{equation}
R:\Delta \longrightarrow S
\end{equation}
that fixes the points of $S$, that is, 
\begin{equation}
R(s)=s
\end{equation}
for all $s \in S$. One can visualize the action of $R$ as projecting $\Delta$ onto $S$, possibly in a geometrically convoluted way, and projecting the points of $S$ onto themselves. Next, one defines a set-valued function
\begin{equation}
F': \Delta \longrightarrow 2^{\Delta}
\end{equation}
as $F'(s)=F(R(s))$; this function is upper-semicontinuous, and its fixed points lie in $S$ and are therefore fixed points of $F$. 
\par
Therefore, whenever it is possible to construct $e$ so that $S$ is convex in some $R^m$, then as per the discussion in the preceding paragraph,  Nash equilibrium in pure strategy quantum games is guaranteed. On the other hand, when $S$ is not convex, for example when $S=e\left(\mathbb{C}P^1\right)=\mathbb{S}^2 \subset \mathbb{R}^3$, then one can extend to its convex hull ${\rm Conv}(S)$ (taking into account the conditions outlined in the Caveats section below) via the convex hull operation, call it $\mathcal{C}$, and note that by Caratheodory's theorem \cite{Cath} ${\rm Conv}(S)$ is compact. It is a well-established topological fact that a non-empty, convex compact subset of $\mathbb{R}^m$ is homeomorphic to the  closed unit ball
\begin{equation}
\mathbb{B}^m=\left\{(x_1,\dots , x_m) \in\mathbb{R}^m : \sum_{j=1}^{m} x_j^2\leq1 \right\}.
\end{equation}
Let $h$ be the homeomorphism ${\rm Conv}(S) \cong \mathbb{B}^m$. The set-valued function $F$ in (\ref{eqn::embed}) now has factors
\begin{equation}
S \xrightarrow{ \mathcal{C}} {\rm conv}(S) \xrightarrow{h} \mathbb{B}^m  \xrightarrow{\mathcal{F}} 2^{\mathbb{B}^m}
\end{equation}
with continuous $\mathcal{F}$.

\begin{figure}
 \begin{tikzcd}[column sep=large, scale=1.2,transform shape]
\otimes_{i=1}^N \mathbb{C}P^{d_i} \arrow[hookrightarrow]{d}{e} \arrow{r}{Q_C} & 2^{(\otimes_{i=1}^N \mathbb{C}P^{d_i})} \arrow{r}{b'}& \otimes_{i=1}^N \mathbb{C}P^{d_i} \\%
S \arrow{d}{h\circ \mathcal{C}} \arrow{rru}[swap]{e^{-1}}
\\
\mathbb{B}^m \arrow{r}{\mathcal{F}} \arrow[bend right]{r}[swap]{\mathcal{F'}} \arrow[bend left]{u}{r} & 2^{\mathbb{B}^m}  \arrow{ul}[swap]{b}
\end{tikzcd}
\caption{Establishing the guarantee of Nash equilibrium in pure quantum strategies in a quantum game $Q$ using Nash's embedding theorem. The functions $b$ and $b'$ take a fixed-point from the power set back to the set.}
\label{fig} 
\end{figure}
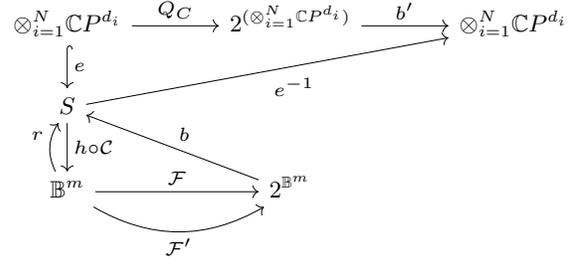
\par
Next, set the payoff function $\mathcal{F}_i=Q_i$ to be any linear function with the property that the elements of its image can be pre-ordered so that(\ref{counteringeq}) and (\ref{linearity}) may be invoked. This produces convex countering sets $\mathcal{F}(h(\mathcal{C}(e(q))))$ of the image $e(q)$ of a play $q$ of $Q$ and guarantees a fixed point for $\mathcal{F}$. To ensure the existence of fixed-points of $\mathcal{F}$ over $S$, construct a retract
\begin{equation}
r: \mathbb{B}^m  \longrightarrow S
\end{equation}
and define the set-valued function
\begin{equation}
\mathcal{F}'(v)=\mathcal{F}(r(v))
\end{equation}
so that the fixed points of $\mathcal{F}'$ lie in $S$ and hence are the fixed points of $\mathcal{F}$. Mapping the fixed points back to $\mathbb{C}P^n$ via $e^{-1}$ gives a Nash equilibrium $q^*$ in $Q$.

This construction is captured in FIG. \ref{fig} for the relevant $n$. Note that when  $S$ is convex, ${\rm Conv}(S)=S$ and our construction restricts to an application of the Kakutani fixed-point theorem. Therefore, up to the caveats mentioned in the following section, our construction guarantees Nash equilibrium in general. Identifying the values of $m$ for which $\mathbb{C}P^n$ embeds as a convex subset into $R^m$ appears to be an open problem to the best of our knowledge.

\section{Caveats}\label{cav}
The Nash equilibrium guarantee due to the construction in FIG. \ref{fig} comes with the conditions that $S$ is non-empty and that $S$ does not equal the boundary of ${\rm Conv}(S)$. We elaborate below. 
\par
Denote by $\partial$ the boundary of a non-empty set. Then, 
$h(\partial {\rm Conv}(S)) \cong \partial h({\rm Conv}(S)) = \partial \mathbb{B}^m = \mathbb{S}^{m-1}$, where 
\begin{equation}
\mathbb{S}^{m-1}= \left\{ (x_1,\dots, x_n) \in \mathbb{R}^n: \sum_{j=1}^m x_j^2 =1\right\}
\end{equation}
is the $(m-1)$-dimensional sphere.  Again, it is well-established that 
no retract exists from $\mathbb{B}^m$ to $\mathbb{S}^{m-1}$ \cite{Kannai} (the no-retract theorem). Therefore, if the Nash embedding $S=\partial {\rm Conv}(S)$, then the construction in FIG. \ref{fig} will fail to guarantee Nash equilibrium. This is certainly true for $N=1$ and $d_i=1$ in the definition of a quantum game in (\ref{quantgame}), giving $\mathbb{C}P^1 = \mathbb{S}^2 =\partial \mathbb{B}^2$ (recall that $\mathbb{C}P^1$ {\it is} $\mathbb{S}^2$ and hence the latter is the natural Nash embedding) as the single player's strategy set. Fortunately, since we are concerned only with multiplayer games, $N \geq 2$.
\par
In general, the elements of any set are the extreme points of its convex hull, that is, points $x$ and $y$ in the convex hull for which $\lambda x+ (1-\lambda)y$ implies that either $\lambda = 0$ or $\lambda =1$. Moreover, the set of extreme points generally forms a proper subset of the boundary of a convex set. Hence, 
\begin{equation}\label{imp}
S \subset \partial {\rm Conv}(S) \implies h(S) \subset \mathbb{S}^{m-1}.
\end{equation}
Next, to see that it is possible to have a retract from $\mathbb{B}^m$ to $h(S) \subset \mathbb{S}^{m-1}$, note that every continuous function from $\mathbb{B}^m$ to $\mathbb{S}^{m-1}$ moves at least one point of the $(m-1)$-sphere by the no-retract theorem. The extreme case where all the points move is undesirable, but any other situation gives a subset $\mathcal{S}$ of fixed points of $\mathbb{S}^{m-1}$ over which a retract from $\mathbb{B}^{m}$ may be constructed. Setting $h(S)=\mathcal{S}$ will give a class of retracts to serve as the function $r$ in FIG. \ref{fig}. For example, the function
\begin{equation}
f: (x_1,\dots, x_m) \rightarrow \left(x_1, \dots, x_{m-1}, \sqrt{1-\sum_{j=1}^{m-1}x_j^2} \right)
\end{equation}
has the "upper hemisphere" as retract.

\section{An application of pure strategy quantum games}  \label{app}
As an application of Nash equilibrium in pure quantum strategies, consider the problem of preparing a $n$ qudit state. Due to decoherence errors, quantum states can deviate from some desired configuration. To model the state preparation and decoherence errors as a non-cooperative pure strategy quantum game, we introduce $N$ notional players so that each player prefers the prepared state to be closest to his desired configuration. 
A state preparing quantum mechanism $Q$ can now be viewed as a non-cooperative, pure strategy quantum game that maps a play of the game to a quantum superposition $q= \sum_{j=1}^n \alpha_j b_j \in \otimes_{i=1}^n\mathbb{C}P^{d_i}$ with $\sum |\alpha_j|^2=1$, and  in which the payoffs are defined by the overlap of the prepared quantum state relative to the desired configuration of each player,  or in other words, the inner-product of the two states. This linear payoff function is  
\begin{equation}\label{linear}
Q_{i}(q) = \langle \psi_i , q \rangle
\end{equation}
with $\psi_{i} \in \otimes_{i=1}^N \mathbb{C}P^{d_i}$ the preferred state of the $i$-th player. The fact that the complex numbers can be pre-ordered, combined with the linearity of $Q_i(q)$, allows one to invoke (\ref{counteringeq}) and (\ref{linearity}) to establish the convexity of $\mathcal{F}(h(\mathcal{C}(e(q))))=C_{e(q)}$. 
\par
By an application of the Nash embedding theorem and the Kakutani fixed-point theorem, we conclude that this game has fixed-point guarantee of Nash equilibrium and that there is an optimal solution to the problem of preparing a pure quantum state under the constraint of decoherence. A similar game model can also be applied to parameter estimation problems, especially in the context of quantum logic gates.\cite{Teklu, Teklu1}.
\par
In contrast, we may also consider the case of a quantum game in which the payoff is defined with respect to a physical observable, as is typically done in quantum game theory literature. A physical observable is represented by a linear Hermitian operator whose eigenstates define possible outcomes. We may define the expectation value of such an operator with respect to a prepared quantum state $q$ as the corresponding payoff. The payoff function $\bar{Q}_i$ calculates the expected value of $q$ to player $i$ via
\begin{equation}\label{obspayoff}
\bar{Q}_i (q) = \sum_{j=1}^{n} e_j|\alpha_j|^2
\end{equation}
where the $e_j$ are real numbers that numerically reflect the preferences of player $i$ over observable states $b_j$ of $\otimes_{i=1}^n\mathbb{C}P^{d_i}$ and $|\alpha_j|^2$ is the probability with which $q$ measures as $b_j$. The payoff function $\bar{Q_i}$ is not linear in general. Hence, the convexity of $C_{e(q)}$ does not follow in general and neither does a fixed-point guarantee for the existence of Nash equilibrium. This result is consistent with the results based on the quantization schemes a la Eisert et al. which define payoffs via (\ref{obspayoff}) and for which pure quantum strategy Nash equilibrium are known to not exist in general.
\par 
For details on how quantum games are realized experimentally, we refer the reader to \cite{Khan3}. For a complete game-theoretic model for designing two qubit quantum computational mechanisms  at Nash equilibrium, we refer the reader to  \cite{Khan, Khan1}.

\section{Future Applications of Quantum Games}
State amplification quantum algorithms like the famous Grover's algorithm \cite{Grover} may be viewed as non-cooperative games between a player and Nature. Consider the quantum state representing the item being searched for as the preferred configuration of the player with control over $x$ qubits, while Nature, with her $y$ qubits, prefers anything but this configuration,  with $x+y=n$. A multiplayer model is also possible. If the players' payoffs are given linearly as in (\ref{linear}), then the algorithm has a Nash equilibrium solution. If the payoffs are computed via (\ref{obspayoff}), then Nash equilibrium may not exist.
\par
With respect to mixed strategies, a natural application of quantum games would be in the areas of quantum communication or stochastic quantum processes where a coalition of players (Alice and Bob) engage in a non-cooperative way with the eavesdropper (Eve) \cite{Williams}. Alice and Bob want to amplify privacy of the communication whereas Eve does not, and in fact may want to decrease it. If the Alice and Bob coalition and Eve try to achieve their respective outcomes via random quantum processes, then Glicksberg's theorem will guarantee a Nash equilibrium. With this guarantee in place, mechanism design methods can be adopted to find an equilibrium. 
\par
An important class of quantum games would be those that study equilibrium behavior of the subset of generalized quantum measurements on finite dimensional systems known as local operations and classical communication (LOCC), a set that is both compact and convex \cite{Chitambar}. Because LOCC is significant in many quantum information processes as the natural class of operations, and given its compact and convex structures, constructing a non-cooperative finite quantum game model for it would be a worthwhile effort.
\par
Adiabatic quantum computing \cite{Farhi, McGeoch,Lucas} can potentially benefit from the quantum game model. An adiabatic quantum computation starts with a system of $n$ qudits in its lowest energy state. A Hamiltonian $H_I$ is constructed that corresponds to this lowest energy state and another Hamiltonian $H_f$ is used to encode an objective function the solution of which is the minimum energy state of $H_f$. Finally, the  actual adiabatic computation occurs as the interpolating Hamiltonian 
\begin{equation}
H\left(s(t)\right)=s(t)H_I+\left(1-s(t)\right)H_f
\end{equation} 
which is expected to adiabatically transform the lowest energy state of $H_I$ to that of $H_f$ as a function of the interpolating path $s(t)$ with respect to time $t$. For large enough $t$ values, adiabaticity holds; on the other hand, $t$ should be much smaller than its corresponding value in classical computational processes for $H(s(t))$ to constitute a worthwhile effort. 
\par
Note that under exponentiation, $H(s(t))$ corresponds to a time-dependent unitary map from $\otimes_{i=1}^{n}\mathbb{C}P^{d_i}$ to itself that we can view as a zero-sum, non-cooperative quantum game. This quantum game has a notional  player I that prefers an element of $\otimes_{i=1}^{n}\mathbb{C}P^{d_i}$ that corresponds to the lowest energy state of $H_f$. Player II or Nature, prefers anything but this element. Players I and II can be given access to any division of qubits to manipulate respectively via pure quantum strategies. Again, if the payoffs to the players are linear, then a Nash equilibrium is guaranteed.
%
%
%

\section{Acknowledgments}
Faisal Shah Khan is indebted to Davide La Torre and Joel Lucero-Bryan for helpful discussion on the topic of fixed-point theorems. This manuscript has been authored by UT-Battelle, LLC under Contract No. DE-AC05-00OR22725 with the U.S. Department of Energy. The United States Government retains and the publisher, by accepting the article for publication, acknowledges that the United States Government retains a non-exclusive, paid-up, irrevocable, worldwide license to publish or reproduce the published form of this manuscript, or allow others to do so, for United States Government purposes. The Department of Energy will provide public access to these results of federally sponsored research in accordance with the DOE Public Access Plan (http://energy.gov/downloads/doe-public-access-plan).


\begin{thebibliography}{99} 

\bibitem{Albers} S. Albers, S. Eilts, E. Even-Dar, Y. Mansour,L. Roditty, {\it On Nash equilibria for a network creation game}, Proceedings of the seventeenth annual ACM-SIAM symposium on discrete algorithm, pages 89-98, 2006.

\bibitem{Fabrikant} A. Fabrikant, A. Luthra, E. Menva, C. Papdimitriou, S. Shenker, {\it On a network creation game}, Proceedings of the twenty-second annual symposium on principles of distributed computing, pages, 347-351, 2003. 

\bibitem{Liu} B. Liu, H. Dai, M. Zhang. {\it Playing distributed two-party quantum games on quantum networks}, Quantum Information Processing (2017) 16: 290. https://doi.org/10.1007/s11128-017-1738-0

\bibitem{Scarpa} G. Scarpa, {\it Network games with quantum strategies}, In: Sergienko A., Pascazio S., Villoresi P. (eds) Quantum Communication and Quantum Networking. QuantumComm 2009. Lecture notes of the Institute for Computer Sciences, Social Informatics and Telecommunications Engineering, vol 36. Springer, Berlin, Heidelberg.

\bibitem{Venkat}  V. Dasari, R. J. Sadlier, R. Prout, B. P. Williams, and T. S. Humble, {\it Programmable multi-node quantum network design and simulation}. In SPIE commercial+ scientific sensing and imaging, pp. 98730B-98730B, (2016). 

\bibitem{Binmore} K. Binmore, {\it Playing for real}, Oxford University Press, 2017.

 \bibitem {Blaquier} A. Blaquiere, {\it Wave mechanics as a two-player game}, Dynamical Systems and Microphysics 33, 1980.
  
\bibitem {Meyer} D. Meyer, {\it Quantum strategies}, Phys. Rev. Lett. 82 (1999) 1052-1055.

\bibitem {Eisert} J. Eisert, M. Wilkens, M. Lewenstien, {\it Quantum games and quantum strategies}, Phys. Rev. Lett. 83, 3077 – Published 11 October 1999.

\bibitem {Nash} J. Nash, {\it  Equilibrium points in N-player games}, Proceedings of the national academy of sciences USA, 36, (1950), 48-49.

\bibitem {Kakutani} S. Kakutani, {\it A generalization of Brouwer's fixed point theorem}, Duke Math. J. Volume 8, Number 3 (1941), 457-459.  

\bibitem {Glick} I. L. Glicksberg, {\it A further generalization of the Kakutani fixed point theorem, with application to Nash equilibrium points}, Proc. Amer. math. soc. 3 (1952) 170–174.  

\bibitem{Nash1} J. Nash, {\it The imbedding problem for Riemannian manifolds}, Annals of Mathematics, 63 (1): 20–63, 1956.

\bibitem{Bengtsson} I. Bengtsson, K. Zyczkowski, {\it Geometry of quantum states: an introduction to quantum entanglement}, Cambridge University Press; 1 edition , anuary 14, 2007.

\bibitem{Brouwer} L. Browuer, {\it Ueber eineindeutige, stetige transformationen von Flächen in sich}, Math. Ann., volume 69, 1910.

\bibitem{Cath} Caratheodory, C. Math. Ann. (1907) 64: 95. https://doi.org/10.1007/BF01449883.

\bibitem{Kannai} Y. Kannai, {\it An elementary proof of the no-retraction theorem},  The American Mathematical Monthly, Volume 88, No. 4, pp. 264-268, 1981.

\bibitem{Teklu} B. Teklu, S. Olivares, M. Paris, {\it Bayesian estimation of one-parameter qubit gates}, Journal of Physics B: Atomic, Molecular, and Optical Physics, 42 (2009) 035502 (6pp).
                      
\bibitem{Teklu1} B. Teklu, M. Genoni, S. Olivares, M. Paris, {\it Phase estimation in the presence of phase diffusion: the qubit case}, Phys. Scr. T140 (2010) 014062 (3pp).

\bibitem{Khan3} F.S. Khan, N. Solmeyer, R. Balu, T.S. Humble, {\it Quantum games: a review of the history, current state, and interpretation}, arXiv:1803.07919 [quant-ph]

\bibitem {Khan1} F.S. Khan, S.J.D. Phoenix, \emph{Mini-maximizing two qubit quantum computations}, Quantum Information Processing, Volume 12, Number 12, 2013.
                                                                           
\bibitem {Khan} F.S. Khan, S.J.D. Phoenix, {\it Gaming the quantum}, Quantum Information \& Computation Volume 13 Issue 3-4, March 2013 Pages 231-244.
                      
\bibitem{Grover} L.K. Grover, {\it A fast quantum mechanical algorithm for database search}, Proceedings, 28th annual ACM symposium on the theory of computing, (May 1996) p. 212.

\bibitem {Williams} B. P. Williams, K. A. Britt, T. S. Humble, {\it Tamper-indicating quantum seal}, Phys. Rev. Applied 5, 014001 – Published 4 January, 2016.

\bibitem{Chitambar} E. Chitambar, D. Leung, L. Mančinska, M. Ozols, A. Winter, {\it Everything you always wanted to know about LOCC (but were afraid to ask)}, Communications in Mathematical Physics, May 2014, Volume 328, Issue 1, pp 303–326.

\bibitem {Farhi} E. Farhi,  J. Goldstone, S. Gutmann, M. Sipser {\it Quantum computation by adiabatic evolution}, pre-print: https://arxiv.org/abs/quant-ph/0001106.

\bibitem {McGeoch} C. C. McGeoch, {\it Adiabatic quantum computation and quantum annealing}, Morgan \& Claypool Publishers series, 2014.

\bibitem {Lucas} A. Lucas, {\it Ising formulations of many NP problems}, Frontiers in Physics, Volume 2, 2014.



\end{thebibliography}
\end{document}